\documentclass[journal, 10pt]{IEEEtran}

\usepackage{multirow}
\usepackage[font=scriptsize,caption=false,labelsep=space]{subfig}
\usepackage{mathrsfs}
\usepackage{graphicx,float,wrapfig,epstopdf,amsmath}
\usepackage[]{algorithmicx}
\usepackage{algpseudocode,algorithm}
\usepackage{balance}
\usepackage{enumitem}
\usepackage{cuted}

\usepackage{mathtools,lipsum}
\usepackage{amsmath}
\usepackage{amssymb}
\usepackage{amsthm}
\usepackage{graphicx}
\usepackage{epstopdf}
\usepackage{times}
\usepackage{textcomp,cite}

\usepackage{url}
\usepackage{hyperref}

\usepackage{multirow}
\usepackage{threeparttable}
\graphicspath{{./figures/}}

\usepackage[table]{xcolor}
\definecolor{infocolor}{RGB}{213,229,255}
\definecolor{inteins}{RGB}{128,179,255}
\definecolor{color1}{RGB}{199,209,232}
\definecolor{color2}{RGB}{230,231,233}

\begin{document}

	\title{Terahertz-Band Integrated Sensing and Communications: Challenges and Opportunities
	}

	\author{\IEEEauthorblockN{Ahmet M. Elbir, \textit{Senior Member, IEEE,}
			Kumar Vijay Mishra, \textit{Senior Member, IEEE,} \\
			Symeon Chatzinotas, \textit{Fellow, IEEE,} 
			and Mehdi Bennis, \textit{Fellow, IEEE}
		}

		\thanks{This work was supported in part by the  the Horizon Project TERRAMETA.
		}
		\thanks{A. M. Elbir is with	the Department of Electrical and Electronics Engineering, Istinye University, 34396 Istanbul, Turkey, and the SnT 
			Interdisciplinary Centre for Security, Reliability and Trust (SnT) at the University of Luxembourg, Luxembourg (e-mail: ahmetmelbir@ieee.org).} 
		
		\thanks{K. V. Mishra is with the United States DEVCOM Army Research Laboratory, Adelphi, MD, 20783, USA; and  with the SnT at the University of Luxembourg, Luxembourg (e-mail: kvm@ieee.org).}
		
		\thanks{S. Chatzinotas is with the SnT at the University of Luxembourg, Luxembourg (email: symeon.chatzinotas@uni.lu). }	
		
		\thanks{M. Bennis is with the Centre for Wireless Communications, the	University of Oulu, Finland (e-mail: mehdi.bennis@oulu.fi). }

	}
	\maketitle
	
	\begin{abstract}

		The sixth generation (6G) wireless networks aim to achieve ultra-high data transmission rates, very low latency and enhanced energy-efficiency. To this end, terahertz (THz) band is one of the key enablers of 6G to meet such requirements. The THz-band systems are also quickly emerging as high-resolution sensing devices because of their ultra-wide bandwidth and very narrow beamwidth. As a means to efficiently utilize spectrum and thereby save cost and power, THz \textit{integrated sensing and communications} (ISAC) paradigm envisages a single integrated hardware platform with a common signaling mechanism.
		However, ISAC at THz-band entails several design challenges such as beam split, range-dependent bandwidth, near-field beamforming, and distinct channel model. This article examines the technologies that have the potential to bring forth ISAC and THz transmission together. In particular, it provides an overview of antenna and array design, hybrid beamforming, integration with reflecting surfaces and data-driven techniques such as machine learning. These systems also provide research opportunities in developing novel methodologies for channel estimation, near-field beam split, waveform design, and beam misalignment.
	\end{abstract}
	%
	

	\section{Introduction}
	\label{sec:Introduciton}
	Lately, the millimeter-wave (mm-Wave) spectrum has been extensively investigated for the fifth-generation (5G) wireless networks to address the demand for high data rates. While the mm-Wave band provides tens of GHz bandwidth, the future sixth-generation (6G) wireless networks are expected to achieve substantial enhancement of data transmission rates ($>100\text{Gb/s}$), low latency ($<1\text{ms}$), and ultra-reliability ($>99.999\%$). In this context, the terahertz (THz) band ($0.1-10$ THz) is expected to be an essential enabling technology in 6G for 2030 and beyond~\cite{thz_Akyildiz2022May}. To this end, the US Federal Communications
	Commission (FCC) has already invited new experimental licenses at $95$ GHz and $3$ THz~\cite{ummimoTareqOverview}.
	
	In addition to the improvement of existing communications technologies in 6G, an unprecedented  paradigm shift is envisioned on the integration of ultra-reliable communications with high-resolution sensing~\cite{elbir2021JointRadarComm}. 
	Further, to save hardware cost and improve resource management, \textit{THz integrated sensing and communications} (ISAC) has been recently suggested to jointly harness the key benefits of THz-band, e.g., ultra-wide bandwidth and enhanced pencil beamforming with common signaling mechanism~\cite{elbir2021JointRadarComm,fanLiuSurveyJSTSP_Zhang2021Sep}. Combining THz communications with THz sensing functionalities finds applications in vehicle-to-everything (V2X), indoor localization, radio-frequency (RF) tagging, and extended/virtual reality.

	Initial ISAC systems had sensing and communications (S\&C) systems operating separate hardware in the same frequency bands and using techniques to avoid interference from each other. However, with increasing convergence between S\&C operations, joint hardware is required. {The ISAC systems, therefore, are broadly classified into radar-communications coexistence (RCC) and dual-functional radar-communications (DFRC)~\cite{fanLiuSurveyJSTSP_Zhang2021Sep}. Specifically, RCC involves both subsystems with their waveforms, which account for the inter-subsystem interference. In contrast, DFRC aims performing both sensing and communication (S\&C) tasks simultaneously via employing a common waveform. The evolution from RCC to DFRC requires the usage of common waveforms, integrated transmit/receive hardware design, and joint processing techniques.} While 
	existing mm-Wave communications protocols/waveforms, e.g.,
	the IEEE 802.11ad standard wi-fi protocol, have been
	proposed for communications-aided vehicular sensing, recent studies have employed similar signaling methods for low THz ($0.06-4$ THz) vehicle-to-vehicle (V2V) ISAC~\cite{thz_jrc_V2X_Petrov2019May}. The
	Third-Generation Partnership Project (3GPP) Release-16 specifies 5G localization and sensing in monostatic mode through time difference-of-arrival (TDoA)~\cite{elbir2021JointRadarComm}. 
	Currently, there exists a work item S1-220144 on ISAC in the 3GPP targeting Release-19.
	
	%
	%

	\begin{figure*}[t]
		\centering
		{\includegraphics[draft=false,width=\textwidth]{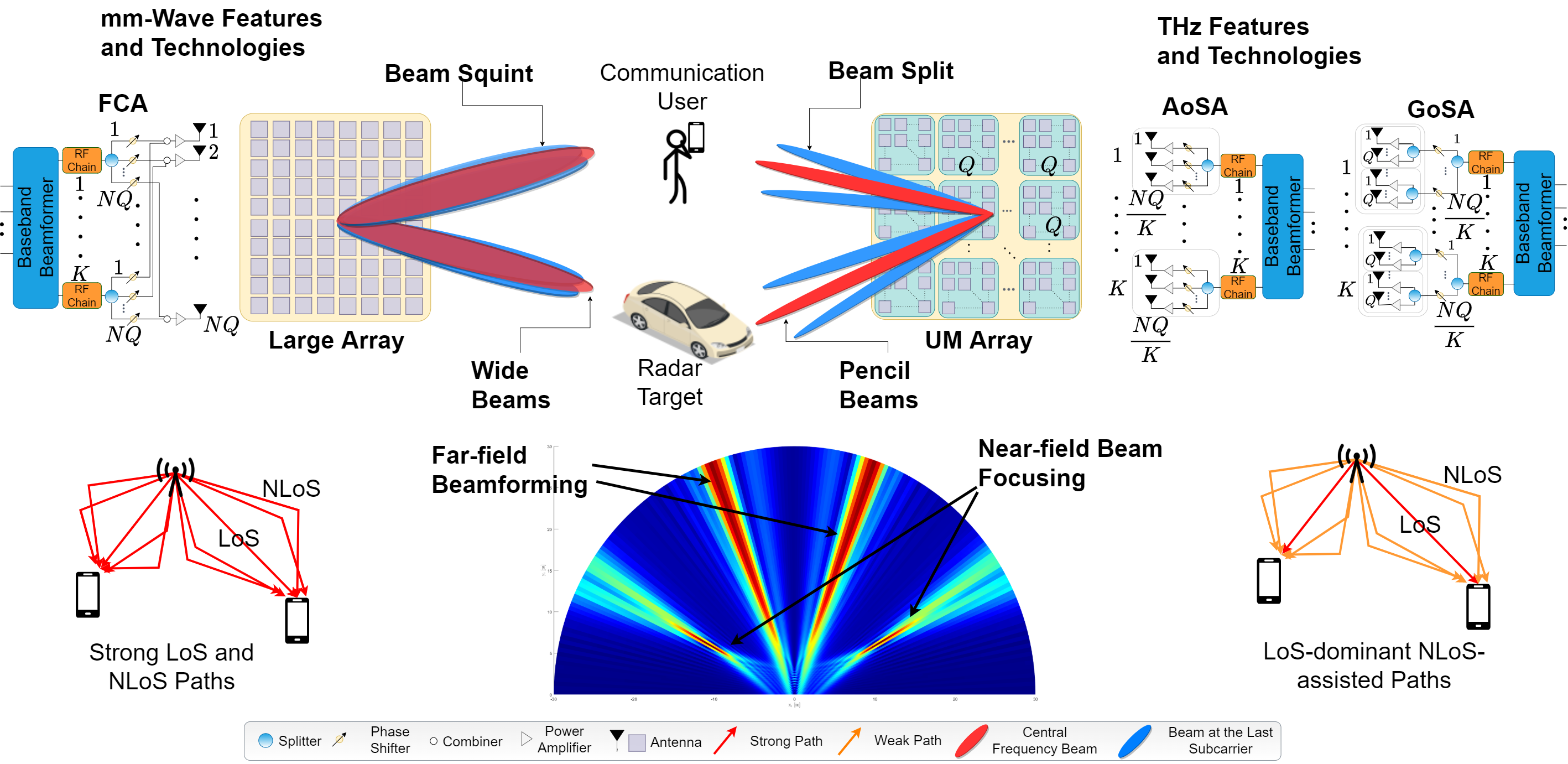} }
		\caption{Comparison of mm-Wave and THz-band characteristics for ISAC  design including multipath components, beam alignment, far/near-field beamforming, and antenna array structures. }
		\label{fig_Diag}
	\end{figure*}

	Certain characteristics of mm-Wave become more aggravated at THz such as high path loss, short transmission range, extreme channel sparsity, and beam squint (Fig.~\ref{fig_Diag}). 
	To overcome these challenges, new signal processing techniques and hardware are required for THz-ISAC design. For instance, analogous to their massive multiple-input multiple-output (MIMO) counterpart in mm-Wave, the  ultra-massive (UM) MIMO configurations are developed to compensate for high path loss in THz~\cite{ummimoTareqOverview}. Further, novel approaches are needed for reliable S\&C performance in terms of channel modeling and wideband signal processing because of THz-specific peculiarities such as beam split, distance-dependent bandwidth, and severe Doppler-induced interference.

	While there exist extensive surveys separately on both THz communications~\cite{thz_Akyildiz2022May,ummimoTareqOverview} and ISAC~\cite{fanLiuSurveyJSTSP_Zhang2021Sep,elbir2021JointRadarComm}, 
	the THz-ISAC remains relatively overlooked. \textcolor{black}{In~\cite{thz_isac1_Han2022Sep}, THz-ISAC is examined from a physical layer perspective with limitation to the channel types and waveform design. In contrast, 	this article examines potential technologies to bring forth these two 6G enablers | THz transmission and ISAC | along with system characteristics/requirements, challenges, and potential solution paths.} In the next section, we		first introduce the unique features of THz-band and their implications, related requirements, and trade-offs for THz-ISAC design. Next, we discuss antenna/array design, hybrid beamforming, integration with intelligent reflecting surfaces (IRSs), and machine learning (ML) to meet THz-ISAC challenges. Finally, we provide a synopsis of research opportunities in THz channel acquisition, near-field beam split, waveform design, beam misalignment, and interference management.

	\section{THz-Band Characteristics for ISAC Design}

	Compared to the mm-Wave channel, the THz channel exhibits certain unique characteristics (Fig.~\ref{fig_Diag}). In what follows, we investigate them along with their implications, requirements and trade-offs for reliable THz-ISAC design.
	
\begin{figure}[h]
		\centering
		{\includegraphics[draft=false,width=\columnwidth]{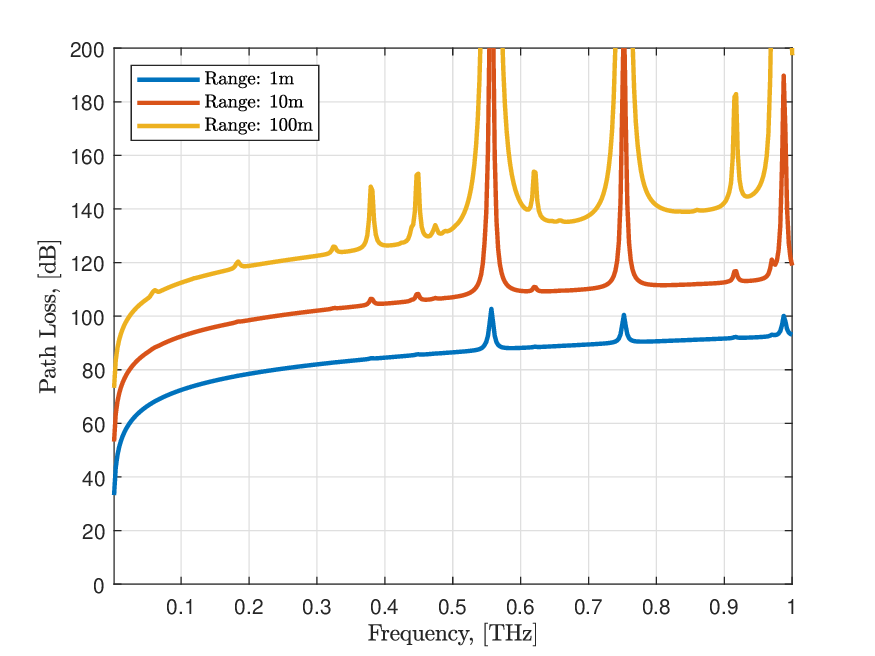} } 
		\caption{Path loss in the THz band due to molecular absorption for various transmission ranges.
		}
		\label{fig_pathloss}
	\end{figure}

	\textit{\textbf{Path Loss:}} The THz channel faces severe path loss ($\sim120$ dB$/100$ m at $0.6$ THz~\cite{delayPhasePrecoding_THz_Dai2022Mar}) governed by both the spreading loss and molecular absorption, which is more significant than the mm-Wave as illustrated in Fig.~\ref{fig_pathloss}~\cite{ummimoTareqOverview}.  In THz-ISAC systems, the radar echo signals experiencing high path loss may still cause stronger interference to the communications system. The high path loss is compensated by beamforming gain through UM antennas that generate multiple beams toward both communications users and radar targets.   {In radar remote sensing applications, surface roughness leads to increased backscatter reflections, resulting in more significant signal loss~\cite{linkbudget_radarPhippen}.}

	\textit{\textbf{Transmission Range:}} {\color{black} A THz-ISAC system has the trade-off that the transmission distance should be long for sensing, while communications tasks may require shorter ranges~\cite{ummimoTareqOverview}.  The valid range for THz radar or communication subsystems is determined by different system parameters such as antenna array size, transmit power, propagation environment, etc. With the decrease of communication delay spread in the THz band, the cyclic prefix (CP) duration can be reduced in the multi-carrier systems. However, with classical orthogonal frequency division multiplexing (OFDM) sensing algorithms, the round-trip delay of sensing targets should be smaller than the CP duration. Thus, the sensing distance might be limited by the CP duration even when the link budget is sufficient~\cite{thz_jrc_OFTS_Wu2022Feb}. } 
	
\begin{figure}[h]
		\centering
		{\includegraphics[draft=false,width=\columnwidth]{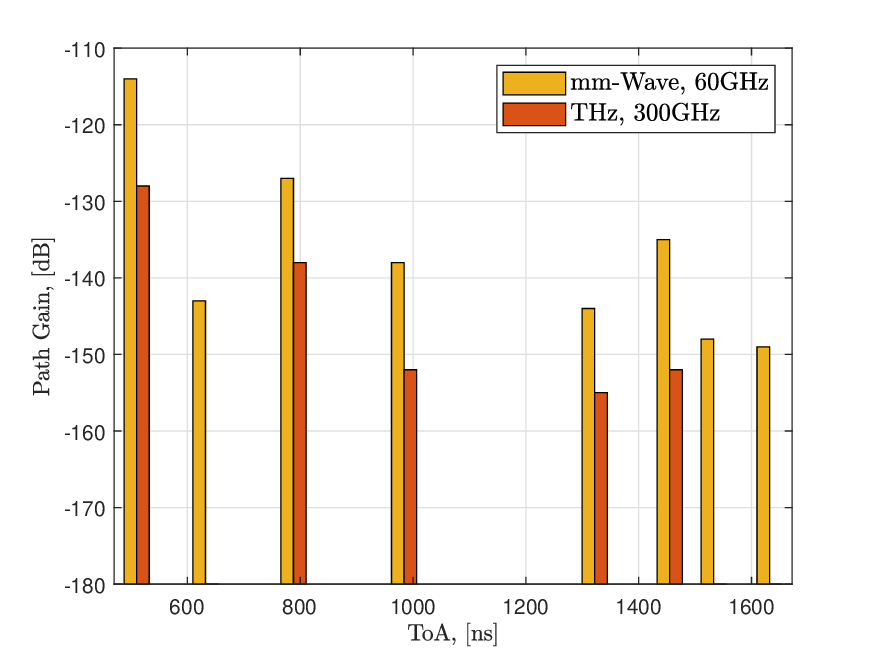} } 
		\caption{Path gain and time-of-arrival  (ToA) for mm-Wave and THz propagation.
		}
		\label{fig_multipaths}
	\end{figure}

	\textit{\textbf{Multipath and Beam Generation:}}  {The THz channel is sparser compared to its mm-Wave counterpart. Specifically, at $300$ GHz, the number of surviving multipath components is up to $5$, while mm-Wave transmission receives about $8$ multipaths, as shown in Fig.\ref{fig_multipaths} \cite{ummimoTareqOverview}. Moreover, the path gain difference between line-of-sight (LoS) and non-LoS (NLoS) paths is $15$ dB higher in THz than in mm-Wave. Thus, the THz channel is characterized as a LoS-dominant and NLoS-assisted model.~\cite{elbir2021JointRadarComm}.} While a THz-ISAC system can benefit from  NLoS paths to improve spatial diversity, especially for 	communications with low-resolution beamformers,  the highly attenuated NLoS links imply fewer secondary echoes in THz sensing/radar applications.		\textcolor{black}{The detection of the NLoS targets is 		an interesting and very challenging issue, for instance when  the moving targets such as 		vehicles, pedestrians and unmanned aerial vehicles (UAVs) may fall into the shadow region of the radar.} \textcolor{black}{The target detection/sensing requires sweeping beams to scan possible targets in the surrounding environment while  communications prefers stable beams toward users to enable tractable data detection~\cite{thz_jrc_waveformDesign_Mao2022Aug}.}

	\begin{table*}
		\caption{State-of-the-art in THz-ISAC Design
		}
		\label{tableSummary}
		\centering
		\begin{tabular}{p{0.17\textwidth}p{0.23\textwidth}p{0.26\textwidth}p{0.24\textwidth}}
			\hline 
			\hline
			\cellcolor{color2}\bf Application &\bf  Signal Processing Techniques\cellcolor{color1} & \cellcolor{color2}\bf Advantages
			& \cellcolor{color1}\bf Drawbacks  \\
			\hline
			\cellcolor{color2}\bf	Hybrid beamforming \cite{elbir2021JointRadarComm}
			& \cellcolor{color1} Joint manifold optimization for the single-user multi-target case
			& \cellcolor{color2} Energy-efficient and corrects beam split without additional hardware
			& \cellcolor{color1} SE degradation due to fewer DoF   \\
			\hline
			\cellcolor{color2}\bf 	IRS-assisted hybrid beamforming   \cite{thz_jrc_beamforming_IRS_Liu2022Mar}
			& \cellcolor{color1} Beampattern generation via Proximal policy optimization
			& \cellcolor{color2} Joint design of transmit and IRS beamformers with enhanced capacity
			& \cellcolor{color1} Only narrowband THz scenario is considered \\
			\hline
			\cellcolor{color2}	\bf OTFS-based waveform design \cite{thz_jrc_OFTS_Wu2022Feb}
			& \cellcolor{color1} DFT-spread OTFS design with superimposed pilot signals
			& \cellcolor{color2} Robustness against the Doppler shift and reduced PAPR compared to OFDM
			& \cellcolor{color1} High receiver cost and complexity \\
			\hline
			\cellcolor{color2}	\bf OFDM-based waveform design \cite{thz_jrc_non_uniformOFDM_Wu2021Apr}
			& \cellcolor{color1} Non-uniform multi-wideband OFDM signaling
			& \cellcolor{color2} Low receiver complexity
			& \cellcolor{color1} Subcarrier spacing depends on Doppler shift \\
			\hline
			\cellcolor{color2}	\bf Beam alignment \cite{thz_jrc_beamAlignment_Chen2022Apr}
			& \cellcolor{color1} Sensing assisted SSB burst transmission
			& \cellcolor{color2} Reduces beam misalignment by $70\%$
			& \cellcolor{color1} Performance depends on sensing capability  \\
			
			\hline
			\hline
		\end{tabular}
	\end{table*}

	\textit{\textbf{Wideband Beam Split:}} The subcarrier-independent analog beamformers largely used in the wideband systems may lead to \textit{beam split} effect in THz channels: the generated beams split into different physical directions at each subcarrier due to ultra-wide bandwidth~\cite{thz_channelEst_beamsplitPatternDetection_L_Dai,delayPhasePrecoding_THz_Dai2022Mar}. This phenomenon has also been called \emph{beam squint} in mm-Wave works~\cite{elbir2021JointRadarComm,thz_Akyildiz2022May}. While both beam squint and beam split pertain to a similar phenomenon, the latter has a more severe degradation of achievable rate in communications. In particular, the main lobes of the array gain corresponding to the lowest and highest subcarrier frequencies do not overlap at THz at all while there is a relatively small deviation in the mm-Wave band (Fig.~\ref{fig_Diag}).
	For sensing, the beam split is approximately $4^\circ$ ($1.4^\circ$) for $0.3 $ THz with  $30$ GHz ($60$ GHz with $2$ GHz) bandwidth, respectively for a broadside target direction-of-arrival (DoA)~\cite{elbir2021JointRadarComm,thz_channelEst_beamsplitPatternDetection_L_Dai}. The compensation of beam split in the THz-ISAC systems requires a hardware trade-off: additional devices (e.g., time-delayer networks) are required for beam split-corrected analog beamformer but they are inessential for digital processing tasks like DoA estimation and channel estimation~\cite{elbir2021JointRadarComm,fanLiuSurveyJSTSP_Zhang2021Sep}.


	\textit{\textbf{Near-field Effect:}}  Due to shorter transmission distance, the THz wave emitted from the transmitter impinging on the receive array may be no longer plane-wave. Hence, the spherical-wave propagation model should be considered for near-field transmission, i.e., when the distance is shorter than the Rayleigh distance, which is proportional to the square of the array aperture (see Fig.~\ref{fig_Diag}). While this distance is $4$ m for an array aperture of $0.1$ m in mm-Wave ($60$ GHz), it becomes approximately $40$ m at $0.6$ THz~\cite{ummimoTareqOverview,elbir2023Jun_nearField_beamSplit_CE_NBAOMP}. This manifests as another degree-of-freedom (DoF) in the range dimension in THz-ISAC design, unlike its mm-Wave counterpart. It may be used to mitigate interference in both angle and distance domains via beam-focusing rather than beam-steering for both sensing targets and communications users. Near-field effects may introduce complex objective functions such as bi-quadratic matrices to the waveform design problem that also necessitates developing low-complexity algorithms.
	


	{
		\textit{\textbf{Distance-dependent Bandwidth:}}  As the transmission distance increases, the THz-specific molecular absorption becomes significant in varying THz-bands, which defines multiple usable transmission windows, each of which is tens of hundreds of GHz wide, and they are separated with absorption peaks, a phenomenon called \textit{broadening of the absorption lines}~\cite{ummimoTareqOverview}. Furthermore, the bandwidth of each of these transmission windows shrinks with the distance. 
		For instance, the transmission window $0.55-0.75$ THz (see Fig.~\ref{fig_pathloss}) may be used entirely for $1$ m range while only $0.6-0.7$ THz of the same band is available for $10$ m range~\cite{elbir2021JointRadarComm,delayPhasePrecoding_THz_Dai2022Mar}.	
		In THz-ISAC design, distance-aware and bandwidth-adaptive modulations/receivers must include the effects of this phenomenon to their advantage.
	
}

	

	\textit{\textbf{Doppler Shift:}}  In wideband THz systems, the Doppler spread may cause significant inter-carrier-interference (ICI), especially in high mobility scenarios~\cite{ummimoTareqOverview}. For instance, the Doppler shift becomes $10$ times larger at $0.3$ THz than that of $30$ GHz. The severe Doppler effect seriously damages the orthogonality among the subcarriers due to ICI, which makes the OFDM challenging~\cite{thz_jrc_OFTS_Wu2022Feb}. The Doppler shift becomes more dominant because of high carrier frequency thereby worsening the false alarms caused by the range sidelobes~\cite{thz_jrc_waveformDesign_Mao2022Aug}.

	\section{Enabling Technologies for THz-ISAC}
	
	The design of THz-ISAC faces several challenging issues. To combat these challenges, herein, we discuss the key enabling technologies  from hardware design and implementation perspectives along with an extensive discussion on the existing state-of-the-art signal processing techniques (see, e.g., Table~\ref{tableSummary})  \textcolor{black}{while \cite{thz_isac1_Han2022Sep} focuses on channel modeling and waveform	design.}

	\subsection{Antenna and Array Design}
	


\begin{figure}[t]
		\centering
		{\includegraphics[draft=false,width=\columnwidth]{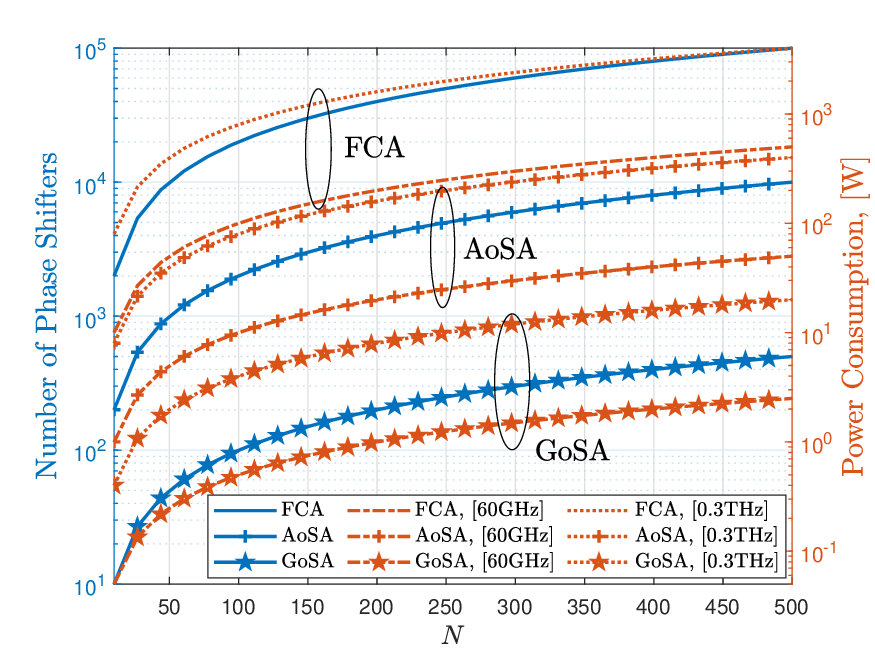} } 
		\caption{(Left) Number of PSs and (Right) power consumption in FCA, AoSA and GoSA architectures, which employ the same number of antennas  $M = NQ$. The number of RF chains is $K = 10$ and $Q = 10$ \cite{elbir2021JointRadarComm}.
		}
		\label{fig_NumPhaseShifters}
	\end{figure}

		To tackle the severe path  loss in THz, 	extremely dense antenna arrays (e.g.,  $5 \times 5$ $ \mathrm{cm}^2$) composed of thousands of antenna elements  are employed~\cite{thz_Akyildiz2022May}. 	Since the number of antennas in THz systems is huge, signal processing with a dedicated RF chain is not efficient even if hybrid analog/digital processing is used. Therefore, subarrayed architectures, e.g., \textcolor{black}{array-of-subarrays (AoSA) and group-of-subarrays (GoSA)~\cite{elbir2021JointRadarComm}} as shown in Fig.~\ref{fig_Diag}, have been proposed for THz S\&C systems as a promising solution against the fully-connected array (FCA) by exploiting the extreme-sparsity of the received THz signal~\cite{elbir2021JointRadarComm}. Consider a THz system with $K$ RF chains and an antenna array with $M = QN$ antennas. Then, the FCA needs $KM$ PSs, whereas AoSA and GoSA employ $QN$ and $N$ PSs, respectively. The main advantage of subarrayed architectures is that they connect a part of the antennas to the same RF chain, thereby reducing the power consumption due to the usage of PSs. Fig.~\ref{fig_NumPhaseShifters} compares these arrays in terms of the number of PSs and  power consumption, which is approximately $5 \mathrm{mW}$ ($40\mathrm{mW}$) at $60$ GHz ($0.3$ THz), respectively~\cite{delayPhasePrecoding_THz_Dai2022Mar}. Here, AoSA and GoSA  exhibit approximately $80$ and $200$ times less consumption compared to FCA. The superiority of GoSA is due to an extra grouping level connecting $Q$ antennas to the RF chain as shown in Fig.~\ref{fig_Diag}.  While the subarrayed connection in AoSA and GoSA enjoys low hardware and energy cost, it yields lower S\&C performance in terms of spectral efficiency (SE) and localization due to fewer DoF than FCA. To address this, overlapped subarrays (OS) are used without additional hardware components. 
	Each of these array setups leads to different S\&C performance as illustrated in Fig.~\ref{fig_HybridBeamforming}) 
	\cite{elbir2021JointRadarComm}. Antenna selection techniques for UM arrays may also be used to yield the best subarray in terms of different communication/sensing performance metrics, e.g., SE, bit-error-rate (BER), and the Cram\'er-Rao lower bound of the target DoAs.
	
	%
	%
	%
	%
	%
	%

	\subsection{Hybrid Beamforming for THz-ISAC}

	Hybrid beamforming is an  enabling technology for THz-ISAC, although it has been mainly introduced for mm-Wave communications systems  to reduce the system cost while providing satisfactory  SE. The hybrid architecture consists of a few digital beamformers and a large number of analog PSs. 	In ISAC, the main aim of hybrid beamforming is to realize a beampattern toward both communications users and radar targets effectively~\cite{fanLiuSurveyJSTSP_Zhang2021Sep,elbir2021JointRadarComm}. 
	The THz-ISAC hybrid beamforming problem faces the following challenges:
	
	\begin{itemize}[wide]
		\item Compared to its mm-Wave counterpart, the THz hybrid beamforming is more challenging due to the UM number of antennas for the solution of the optimization problem, which is highly non-linear and  non-convex due to the coupling between analog/digital beamformers, and the constant-modulus constraint for realizing PSs. 
		\item THz-ISAC hybrid beamforming design should consider THz-specific peculiarities such as  the beam split phenomenon and beam misalignment due to the generation of very narrow beams in THz. Furthermore, the path loss in THz is distance-dependent for which the THz-ISAC system should employ multiple transmission windows for long- and short-distance targets/users.
	\end{itemize}
	
	%

\begin{figure}[t]
		\centering
		{\includegraphics[draft=false,width=\columnwidth]{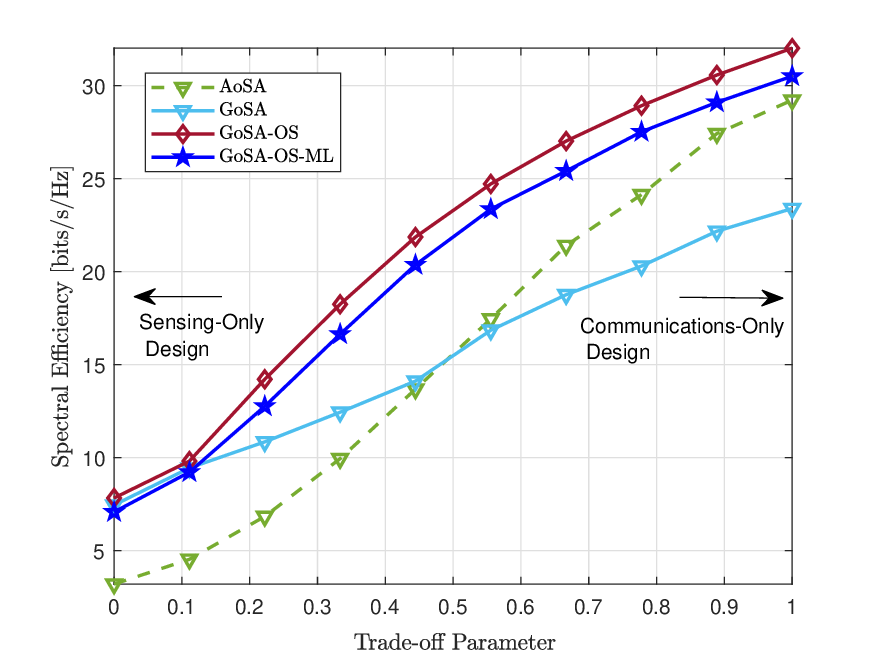} } 
		\caption{SE trade-off in THz-ISAC with respect to hybrid beamformer tuning parameter for AoSA, GoSA and GoSA-OS as well as ML-aided GoSA-OS \cite{elbir2021JointRadarComm}. 
		}
		\label{fig_HybridBeamforming}
	\end{figure}

	Considering the aforementioned challenges, the design of THz-ISAC hybrid beamforming also requires the combination of different performance metrics of sensing (mean-squared-error (MSE) of DoA {and range} estimation) and communications (SE). One possible approach is the optimization of the hybrid beamforming weights jointly with radar- and communications-only beamformers with a tuning parameter~\cite{elbir2021JointRadarComm,fanLiuSurveyJSTSP_Zhang2021Sep}. Herein, the radar beamformer consists of the steering vectors corresponding to the {far-field} target DoAs whereas the communications beamformer is constructed from the singular value decomposition (SVD) of the channel matrix. The tuning parameter controls the trade-off between the accuracy/prominence of S\&C tasks. For instance, as illustrated in Fig.~\ref{fig_HybridBeamforming}, this tuning parameter is usually selected between $0$ (sensing-only design) and $1$ (communications-only design) to optimize the balance over the performance metrics related to both radar and communications.
	
\begin{figure}[t]
		\centering
		{\includegraphics[draft=false,width=\columnwidth]{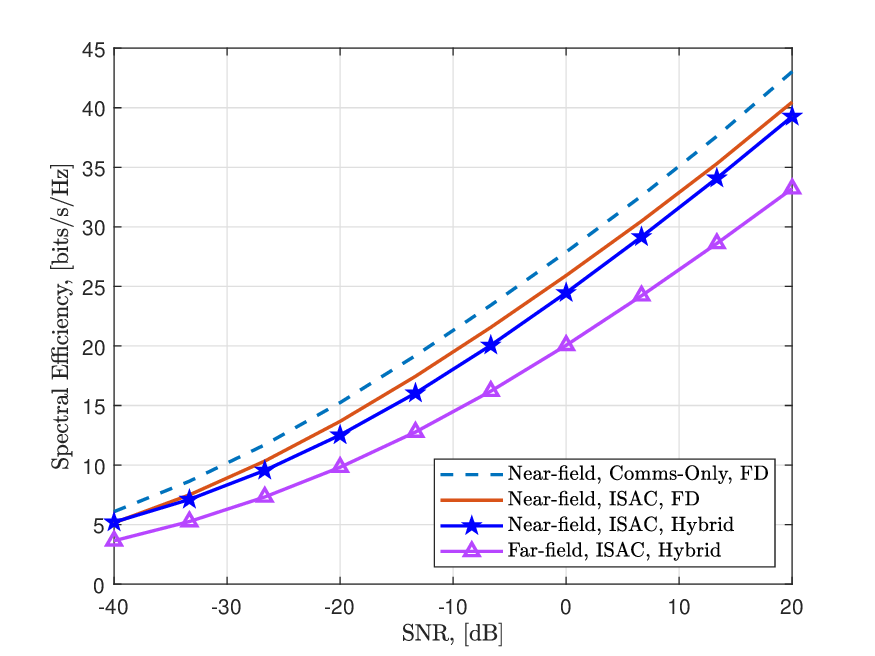} } 
		\caption{Near-field THz-ISAC beamforming performance in terms of SE for FD and hybrid beamforming in communications-only and ISAC scenario \cite{elbir2023Jun_nearField_beamSplit_CE_NBAOMP}.
		}
		\label{fig_NF_BF}
	\end{figure}

	Fig.~\ref{fig_NF_BF} shows the orthogonal matching pursuit (OMP)-based THz-ISAC hybrid beamforming performance, wherein the targets and the users are in the near-field of the BS. We observe that the near-field ISAC beamformer performs close to fully digital (FD) ISAC and communications-only beamformers with the trade-off parameters of 0.5 and 1, respectively. Fig.~\ref{fig_NF_BF} also demonstrates poor SE performance when a far-field assumption is imposed while designing the OMP dictionary.
	
	Due to the usage of subcarrier-independent analog beamformers, the generated beams at central and low-/high-end subcarrier frequencies face a severe array gain loss causing beams to split into different directions (Fig.~\ref{fig_Diag}). One approach to mitigate beam split in THz transmission is realizing the analog beamformer with PSs and time delayers, hence called delay-phase precoding (DPP)~\cite{delayPhasePrecoding_THz_Dai2022Mar}. This approach first generates a subcarrier-independent beamformer, then constructs virtual subcarrier-dependent beams with beam split compensation by using time delayers. The additional time delayer network is expensive because each \textcolor{black}{phase shifter (PS)} should connect multiple time delayers, each of which consumes approximately $100$ mW, which is  more than that of a PS ($40$ mW) in THz~\cite{delayPhasePrecoding_THz_Dai2022Mar}.
	
	The effect of beam split can also be mitigated via signal processing techniques without additional hardware. For instance, \cite{elbir2021JointRadarComm} devises a beam split correction technique, wherein the corruptions in subcarrier-independent analog beamformer due to beam split are computed and passed into subcarrier-dependent digital beamformers which are then corrected to realize beam split-free beampattern. As a result, the high power consumption of time delayer networks requires better signal processing approaches for beam split mitigation.  
	 In addition, developing a common performance metric for S\&C to better represent the system requirements can improve the performance and lower the hardware cost. 
	

	\subsection{IRS-assisted Systems}
	An IRS is a two-dimensional (2D) surface composed of a large number of meta-material elements, reflecting the incoming signal toward the intended direction by introducing a pre-determined phase shift. Thus, the IRS provides improved energy and spectral efficiency in wireless networks. The usage of IRS can be especially advantageous in THz-ISAC in compensating for the high path loss and improving the sensing coverage and communications performance. 
	Compared to conventional ISAC, the IRS-assisted case is more challenging since it involves the joint design of transmitter beamformers and IRS phase shifts. For this purpose, a proximal policy optimization (PPO) approach with reinforcement learning (RL) is proposed in~\cite{thz_jrc_beamforming_IRS_Liu2022Mar}, wherein the transmitter and IRS parameters are jointly optimized for THz-ISAC, wherein the users are also designated as radar targets. However,~\cite{thz_jrc_beamforming_IRS_Liu2022Mar} considers only narrowband scenarios without exploiting the key advantage of ultra-wide bandwidths in THz. {\color{black}Moreover, compared to a single IRS, the use of double IRS improves the radar sensing (communications) signal-to-interference-plus-noise ratio (SINR) performance. Fig.~\ref{figRadarCommSINR} shows the radar and minimal communication  with respect to the noise power. It can be seen that the use of double IRS provides about $4$ dB extra gain for both radar sensing and communication SINR.}
	
	In fact, the IRS-assisted ISAC design is a new paradigm even for the mm-Wave band as it is envisioned for 6G wireless networks. Therefore, several design challenges in IRS-assisted ISAC are unexamined such as wideband processing and waveform design, clutter/multi-user interference suppression, and physical layer security. Besides the conventional IRS, the simultaneously transmitting and reflecting intelligent surface assisted ISAC provides full-space coverage and more DoF, hence, opening new research opportunities. 
	
%
%
%
%
%

	{
	\subsection{UAV-Borne Systems}
	The use of UAVs in THz-ISAC scenarios has garnered significant interest due to their strong air-ground LoS channels. Similar to the IRS-assisted scenario, UAVs can enhance the coverage of S\&C with greater flexibility. However, implementing UAV-assisted ISAC presents several design challenges, such as waveform design and resource allocation (arising from limited battery and flight times). Specifically, UAVs can sense targets in the environment by periodically transmitting probing beams while maintaining communications links with users~\cite{uav_isac_2_Chang2022Mar}. This creates a fundamental trade-off in resource allocation between achieving high beamforming gain and reliability for sensing targets and providing satisfactory communication throughput and connectivity. UAVs equipped with antenna arrays can generate both sensing and communication beams ~\cite{uav_isac_Chen2020Dec}, where an integrated scheduling approach is proposed to manage resource allocation among sensing, communications, and motion control signals.

	%

	%
	\begin{figure}[t]
		\centering
		{\includegraphics[draft=false,width=.9\columnwidth]{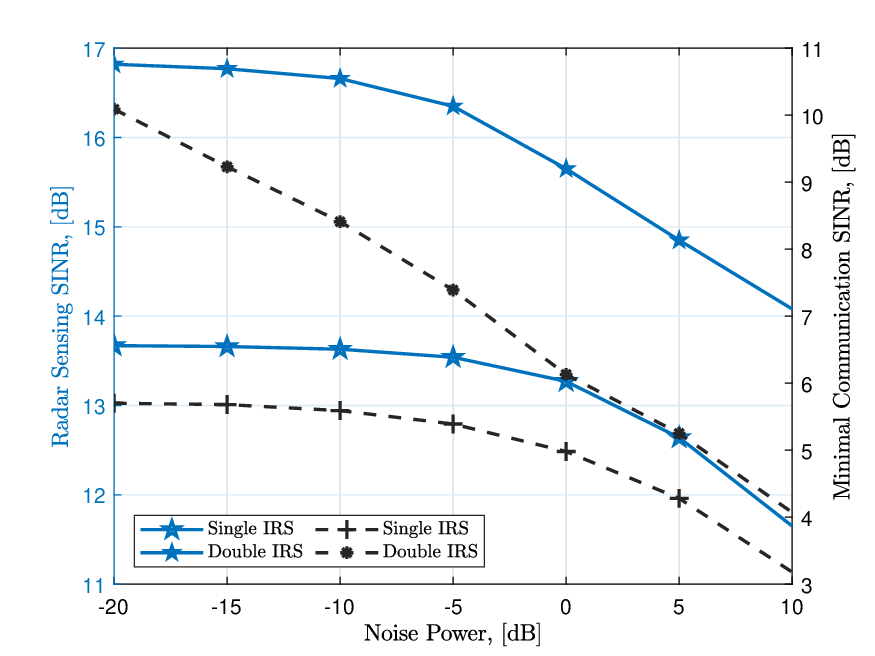} }
		\caption{\color{black}Communication and radar sensing SINR versus noise power. } \vspace{-10pt}
		\label{figRadarCommSINR}
	\end{figure}

	\subsection{ML Solutions for THz-ISAC Design}
	Compared to model-based techniques relying on accurate mathematical expressions, ML-based approaches exhibit three main advantages: robustness against imperfections in the data, environment-adaptivity via retraining with new data and post-training computational complexity. As a result, ML has been regarded as one of the key enabling technologies for 6G wireless networks~\cite{ummimoTareqOverview,elbir2023Jun_nearField_beamSplit_CE_NBAOMP}.

	With the aforementioned benefits, ML-based techniques gained much interest separately for sensing (DoA estimation, localization) and communications (channel estimation, beamforming, resource management) applications. For ML-based THz-ISAC design, one should consider jointly solving multiple problems related to S\&C based on the available training data. For instance, \cite{thz_jrc_beamforming_IRS_Liu2022Mar} devises a reinforcement learning (RL) approach for joint beamformer design at the DFRC and IRS. Also, ML-based THz-ISAC hybrid beamforming is proposed  in~\cite{elbir2021JointRadarComm}, wherein two different learning models (one for DoA estimation and another for beamforming) are designed. This approach achieves approximately $200$ times lower computation time while providing satisfactory SE compared to fully-digital beamforming (see Fig.~\ref{fig_HybridBeamforming}).
	
	\begin{figure}[t]
		\centering
		{\includegraphics[draft=false,width=\columnwidth]{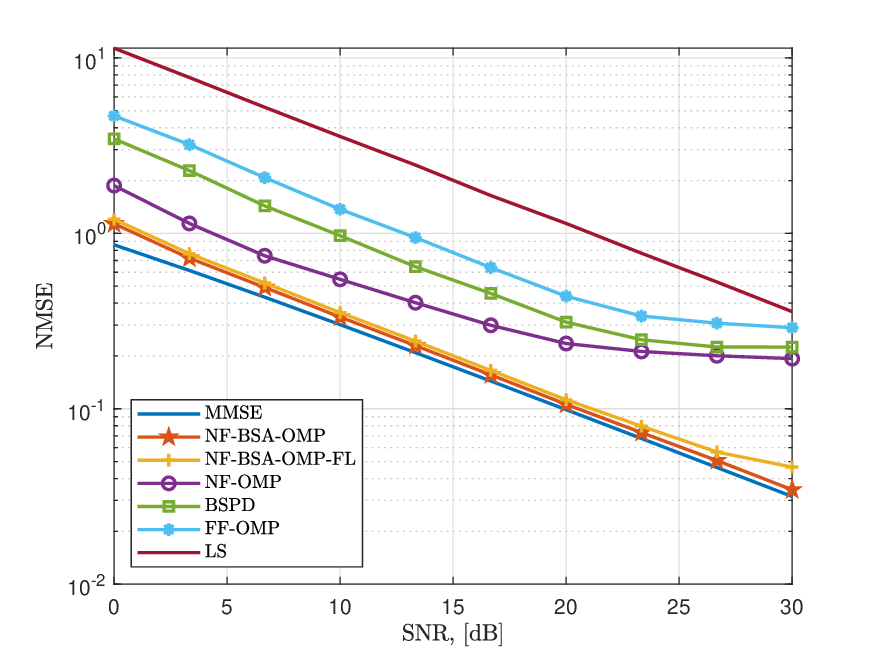} } 
		\caption{Near-field THz channel estimation performance in terms of NMSE for signal processing  (BSA-OMP, BSPD, and LS) as well as  ML-empowered  (BSA-OMP-FL) approaches~\cite{elbir2023Jun_nearField_beamSplit_CE_NBAOMP}.
		}
		\label{fig_ChannelEst}
	\end{figure}

	\begin{figure}[t]
		\centering
		{\includegraphics[draft=false,width=\columnwidth]{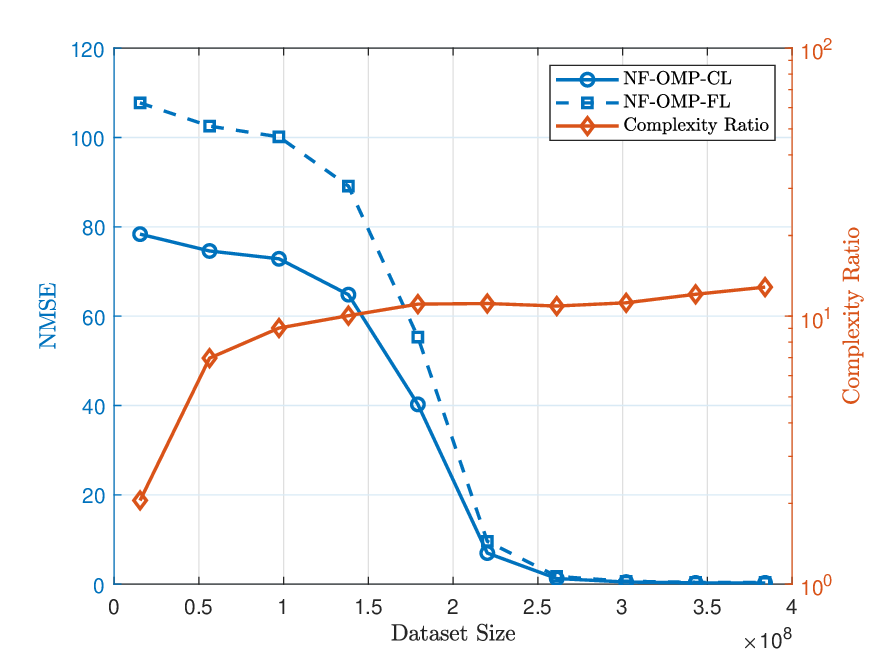} } 
		\caption{CL- and FL-based near-field THz channel estimation versus dataset size, and computational complexity ration of CL and FL.
		}
		\label{fig_FL_complexity}
	\end{figure}

	Most of the ML algorithms rely on collecting the data from the edge devices, e.g., mobile phones, to a central server, wherein the learning model is trained. The size of the datasets usually scales with the number of antennas in the array. Hence, dataset transmission entails huge communications overhead (CO) in centralized learning (CL) schemes. 	To reduce the high CO in CL, federated learning (FL) approach is introduced for near-field THz channel estimation problem~\cite{elbir2023Jun_nearField_beamSplit_CE_NBAOMP} where approximately $12$ times lower CO is obtained while providing satisfactory NMSE performance (see Fig.~\ref{fig_ChannelEst}).  In order to achieve more communications-efficient learning capability, the sparsity of THz channels can be exploited to reduce the size of learning models, thereby developing quantized or compressed neural networks. {Fig.~\ref{fig_FL_complexity} shows the NMSE performance of CL- and FL-based training for near-field THz channel estimation in terms of dataset size. We see that both schemes attain similar NMSE as the learning model is fed with large datasets. We also observe that CL involves approximately $10$ times higher CO for a reasonably large datasets. While FL is efficient in terms of CO, it requires computational hardware, e.g., parallel processing units, at the users to perform model computation. In order to provide an effective solution to optimize CO as well as power consumption at the edge users, hybrid FL-CL approaches are devised as an effective solution~\cite{elbir_hfcl_Elbir2022Jun}, wherein only the users with adequate computational power perform FL while the remaining ones transmit their dataset to the server which computes the model parameters on behalf of them.         }
	
	%
	%


	\section{Open Problems and Research Opportunities}
	The vision of THz-ISAC faces several design challenges, to name some, THz channel characteristics due to shorter S\&C range, waveform design providing a joint signaling as well as beam misalignment and link failures due to pencil beamforming. In the following, we provide an extensive discussion and highlight the related research opportunities \textcolor{black}{on channel acquisition, waveform design, beam alignment, interference and index modulation.}

	\subsection{THz Channel Acquisition}
	
	Compared to its mm-Wave counterpart, THz channel estimation is more challenging due to the involvement of additional error sources to be modeled, e.g., beam split effect, near-/far-field channel modeling, etc. 
	\subsubsection{Beam Split} True-time-delay (TTD) processing is conventionally used for THz channel estimation in the presence of beam split, wherein a time delayer network is used to realize analog beamformers similar to the DPP approach in~\cite{delayPhasePrecoding_THz_Dai2022Mar}. While this approach necessitates additional time delayer network, signal processing-based approaches, e.g., beam split pattern detection (BPSD)~\cite{thz_channelEst_beamsplitPatternDetection_L_Dai} and beam split aware  (BSA) OMP~\cite{elbir2023Jun_nearField_beamSplit_CE_NBAOMP} can also be used for accurate THz channel estimation. {Beam-split is also a challenge in IRS design and optimization. While beam-split can be compensated at base station (BS) via time delayer components, it is challenging and costly to perform it at the IRS as it is a passive device. Instead, the beam-split-corrupted signal can be collected at the BS during beam training process and compensated similar to array calibration process~\cite{elbir2021JointRadarComm,elbir2023Jun_nearField_beamSplit_CE_NBAOMP}.    }
	
	\subsubsection{Near-field Beam Split} In contrast to far-field, wideband THz transmission in near-field causes beams to split in different directions as well as different distances.  This leads to a new phenomenon called \textit{near-field beam split} which is range-dependent and not easily mitigated by direct application of the far-field beam split correction techniques~\cite{elbir2023Jun_nearField_beamSplit_CE_NBAOMP}. The near-field beam split leads to serious problems in both S\&C. In this case, the transmitted signal fails to focus on the desired user/target location, at which only the beams generated at the central frequency can arrive. Furthermore, the radar receiver should take into account designing the matched filters with  the impulse response of the range-dependent propagation channel as well as beam split.	Fig.~\ref{fig_ChannelEst} shows the NMSE performance for near-field wideband THz channel estimation at $300$ GHz with $30$ GHz bandwidth~\cite{elbir2023Jun_nearField_beamSplit_CE_NBAOMP}. Fig.~\ref{fig_ChannelEst} indicates that BSA-OMP attains close to  minimum mean-squared-error (MMSE) estimation performance. We observe that the direct application of the far-field model (FF-OMP and BSPD) as well as the techniques that do not take into account the impact of beam split (NF-OMP and least-squares (LS)) yield poor NMSE performance. The BSA-OMP approach also involves FL which leads to performance loss arising from decentralized model training. 


	\subsection{Waveform Design}
	The ISAC receiver is responsible for accurately  demodulating the received communications signal while recovering the echo signal from the targets. Unlike   ISAC aims to improve the \textit{integration gain} via joint processing of S\&C signals, thereby reducing the hardware requirements~\cite{elbir2021JointRadarComm,fanLiuSurveyJSTSP_Zhang2021Sep}. 
	
	For waveform design, the ISAC resource allocation can be performed in either communications-centric (CC), sensing-centric (SC), or unified design schemes. The former techniques may be easier at the cost of low efficiency; while the latter has improved accuracy in both S\&C with high signal processing and computational complexity. {For instance, a common hybrid analog/digital beamformer is designed in~\cite{elbir2021JointRadarComm} to simultaneously generate beams towards both radar targets and communication users. A more communication-centric design is also considered in~\cite{thz_SPIM_ISAC_Elbir2023Mar}, wherein the spatial paths for the communication links are exploited for index modulation.      }

	\subsubsection{Physical Layer THz-ISAC Waveform Design} The wideband processing is critical in THz-bands, at which the Doppler spread causes ICI, which makes OFDM inapplicable, especially in high mobility cases. To this end, orthogonal time-frequency-space (OTFS) multiplexing techniques can provide robustness against the Doppler shift in THz-ISAC~\cite{thz_jrc_OFTS_Wu2022Feb}. The OTFS is also advantageous in reducing the peak-to-average power ratio (PAPR), from which the OFDM systems suffer. Nevertheless, the OTFS-based waveform design comes with a non-negligible receiver cost and complexity.  Instead, ML-based modulation classification techniques may be more efficient, wherein the 2D time-frequency frames can be used as input data. {\color{black}Also in~\cite{thz_jrc_waveformDesign_Mao2022Aug}, a communication data-embedded multi-band ISAC waveform is proposed for simultaneous high-resolution sensing and communication with low mutual interference.} An efficient design is introduced in~\cite{thz_jrc_non_uniformOFDM_Wu2021Apr} by exploiting non-uniform  multi-wideband (NU-MW) OFDM subcarriers. However, one should design the subcarrier spacing carefully in this technique since it needs to be less than the maximum Doppler shift, which may be application-dependent. In practice, adaptive methods may be helpful for controlling the subcarrier spacing.
	
	The ISAC waveform design techniques should also take into account the THz transmission windows, which may shrink with the transmission range (see Fig.~\ref{fig_Diag}).	By exploiting the transmission windows in THz, distance-aware approaches can be deployed for the THz-ISAC applications. That is, the central part of the bandwidth is dedicated to the long-distance users/targets while the S\&C operations can benefit the whole bandwidth for the short-distance users/targets~\cite{ummimoTareqOverview}.
	
	\subsubsection{\color{black}Waveform Design for THz-ISAC Multiple Access}Most of the ISAC literature concentrates on physical layer design, while there are a few works on higher layer coordination of S\&C with multiple access technologies. For instance, by utilizing the preamble sequence in IEEE 802.11ad frame, a radar-aware carrier-sense multiple access (RA-CSMA) is proposed in~\cite{thz_jrc_V2X_Petrov2019May} for low-THz ($0.06-4$ THz) V2V ISAC. In particular, 	the sensing signals are treated as a packet in CSMA. Although this approach is advantageous in terms of SE, it may lead to a low sensing duty cycle in case of congestion of radars. 

	\subsection{Beam Misalignment} Another challenging issue in THz-ISAC is beam misalignment due to pencil beamforming. In THz, the beamwidth is very narrow such that the beams at the transmitter and the users may not be aligned. While pencil beamforming with narrow beamwidth reduces the randomness of the path loss in the THz-band, it causes link failures, inter-cell handovers, and intra-cell beam switches. An ISAC-like approach (i.e., sensing-assisted communication)  is developed in~\cite{thz_jrc_beamAlignment_Chen2022Apr}, wherein multiple synchronization signal blocks (SSBs) are transmitted to mitigate beam misalignment. It is reported in~\cite{thz_jrc_beamAlignment_Chen2022Apr} that the sensing-aided approach reduces beam misalignment probability by up to $70\%$. On the other hand, the algorithm performance directly depends on the sensing accuracy. {Beam misalignment is also investigated in~\cite{uav_isac_2_Chang2022Mar} for UAV-assisted THz-ISAC scenario, and it is shown that beam misalignment causes significant errors in LoS THz ($300$GHz) channels compared to LoS and NLoS mmWave ($60$GHz) channels due to severe path loss.     }

	\subsection{Interference Management}
	The specific features of the THz-band, such as path loss and the Doppler shift, aggravate the ISAC interference management. 
	The communications systems also suffer from multi-user interference. Similarly, suppression of clutter echoes (reflections from unwanted targets) is always a concern in radar processing. {On the other hand, the use of large antenna arrays makes the transceiver design relatively robust against the interference because the beamwidth is very narrow in THz.}  IRS-aided systems have shown encouraging results for clutter suppression by utilizing echoes from multiple NLoS paths~\cite{thz_jrc_beamforming_IRS_Liu2022Mar,elbir2021JointRadarComm,fanLiuSurveyJSTSP_Zhang2021Sep}.
	Advanced waveform design techniques may be employed such that the S\&C systems ensure multiple access signaling via time, frequency, spatial, and code domains to prevent mutual interference. These non-overlapping resource allocation methods have a rich heritage of research and implementation. However, the objective of ISAC design is to integrate S\&C seamlessly. To this end, recent overlapping resource management techniques have been shown to achieve a unified waveform design by maximizing the  SINR of both systems with increased DoFs~\cite{thz_jrc_OFTS_Wu2022Feb}. 
	
	{\color{black}
		\subsection{Index Modulation}
		In the last decade, index modulation (IM) has been introduced to achieve improved SE than conventional modulation schemes. In IM, the transmitter encodes additional information in the indices of the transmission media such as subcarriers, antennas, spatial paths, etc. Thanks to transmitting additional information bits via the indices of activated RF components (e.g., antennas, subcarriers or beamformers), IM-aided ISAC have shown significant performance improvement. In	particular, the IM-aided ISAC hybrid beamforming exhibit higher SE than that	the use of fully digital beamformers as shown in	Fig.~\ref{fig_SE_SNR} wherein the SE of the THz-ISAC system is evaluated~\cite{thz_SPIM_ISAC_Elbir2023Mar}.
		
		\begin{figure}[t]
			\centering
			{\includegraphics[draft=false,width=\columnwidth]{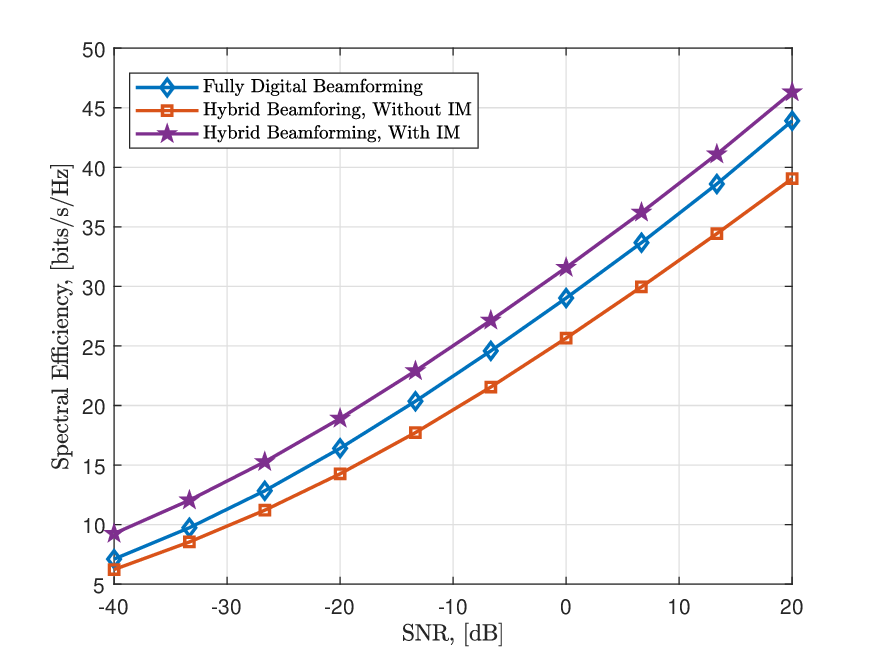} } 
			\caption{IM-aided THz-ISAC hybrid beamforming performance in terms of SE \cite{thz_SPIM_ISAC_Elbir2023Mar}.
			}
			
			\label{fig_SE_SNR}
		\end{figure}

	}
	
	{
	\subsection*{Hardware Design}
	While ISAC aims to reduce hardware costs by utilizing a common platform for both S\&C tasks, hardware design is \textit{a fortiori} challenging at THz frequencies. To compensate for high path loss and accurately receive THz signals, a very large number of antenna elements must be employed to improve beamforming gain. This complexity in THz-ISAC hardware presents a challenge for processing high-dimensional array data for S\&C. One approach is to exploit the sparsity of THz channel data. For instance, the AoSA and GoSA architectures (see Figs. 1 and 3) take advantage of channel sparsity to reduce computational and hardware complexity with fewer phase shifters\cite{ummimoTareqOverview, elbir2021JointRadarComm}. Using the same hardware but implementing advanced signal processing techniques can also enhance S\&C performance. For example, a non-uniform OFDM signaling approach in \cite{thz_jrc_non_uniformOFDM_Wu2021Apr} achieves sub-millimeter-level accuracy for range estimation of radar targets while maintaining satisfactory communications performance at a $100$ Gbps rate at $300$ GHz. Additionally, IM-based approaches have shown improvements in communications rates by transmitting additional information bits via the indices of RF components without adding hardware components or interfering with sensing performance\cite{thz_SPIM_ISAC_Elbir2023Mar}.
	
	To address the severe path loss at THz frequencies, where the wavelength is very small, extremely dense antenna arrays with thousands of elements are employed~\cite{thz_Akyildiz2022May}. Tunable graphene-based plasmonic nano-antennas or metamaterials are used to provide dynamic THz beamforming capability (Fig.\ref{fig_Antennas}) \cite{ummimoTareqOverview, thz_Akyildiz2022May}. The graphene-based structure allows for steering the main-lobe direction by changing the energy levels of the graphene layer. Leaky-wave antennas (Fig.~\ref{fig_Antennas}) have also gained interest for THz applications due to the coupling of frequency and beam angle, making them suitable for THz sensing and tracking applications.
	
	The IRS design is another challenging issue for IRS-assisted THz-ISAC. With the abundant bandwidth available at the THz band, the IRS must account for multiple frequency bands to be used in multi-cell, multi-band systems for various service providers. The phase shifts introduced by the IRS elements must be determined for different frequencies, typically by tuning the capacitor in the IRS elements~\cite{thz_jrc_beamforming_IRS_Liu2022Mar}. Similar to antenna design, graphene-based IRS elements offer high conductivity and electrical configurability, and plasmonic-based antenna patches can be deployed on an IRS for THz applications~\cite{ummimoTareqOverview}.

	}

	\begin{figure}[t]
		\centering
		{\includegraphics[draft=false,width=\columnwidth]{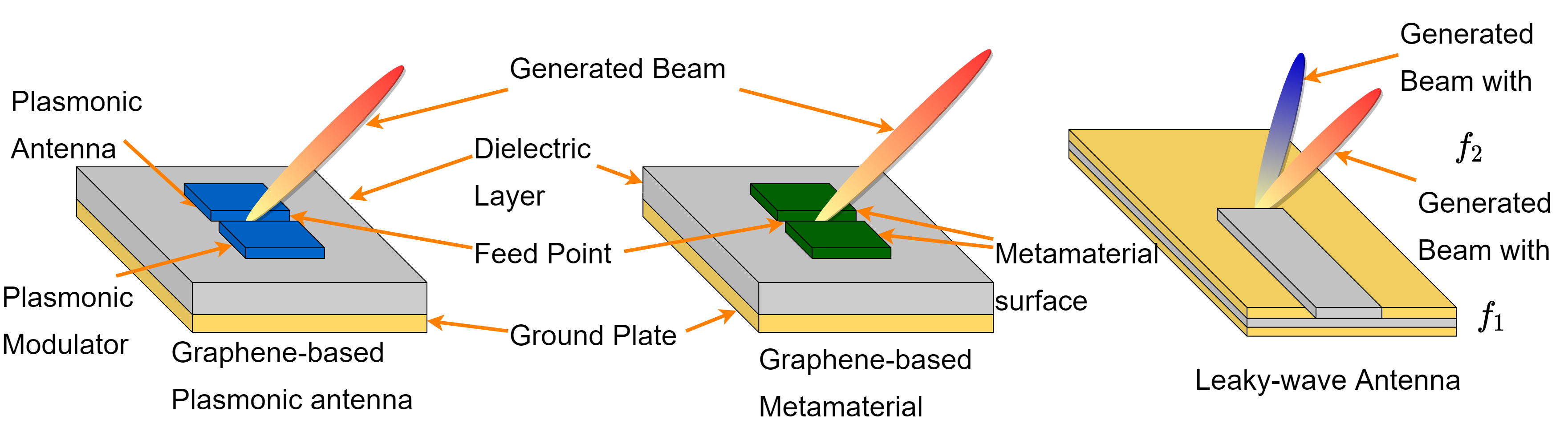} }
		\caption{Antenna designs for THz S\&C: (Left) graphene-based plasmonic antenna and (Middle) metamaterials, and (Right) leaky-wave antenna. }
		\label{fig_Antennas}
	\end{figure}
	
	{
	\subsection{Link Budget}
	Because of high path loss and the reflection/scattering loses, the received signal for both communication and sensing links become very weak (see Fig.~\ref{fig_pathloss}). Therefore, a link budget calculation is helpful to accurately assess the feasibility of THz-ISAC.	The evaluation of link budget involves the calculation of the received SNR at the receiver given the output power at the transmitter. For communications, a link budget analysis of THz systems with $512\times32$ transceiver array is investigated in~\cite{linkBudget2_Rikkinen2020Nov} for a  carrier frequency of $300$GHz with bandwidth of $30$GHz. It is reported that the Tx antenna gain of $26$dBi can be achieved at about $10$m with $3.6^\circ$ beamwidth. For sensing, the authors of \cite{linkbudget_radarPhippen} have shown that a pedestrian (car) are detectable at the ranges of up to $20$m ($50$m) for a THz automotive radar at $290$GHz. While link budget calculation has been extensively studied for communications and sensing purposes~\cite{linkBudget2_Rikkinen2020Nov,linkbudget_radarPhippen}, its evaluation for ISAC is still an open area of research. 
	}

	{
	\section{Summary and Future Outlook}
	We examined THz-ISAC as one of the key enabling technologies of 6G wireless networks. It is a clear outcome of this article that climbing up to THz frequencies will enable high resolution sensing capabilities, which will enforce THz-ISAC applications in 6G for 2030 and beyond. 	We note the following future research directions:
	
	\begin{itemize}[wide]
		\item In terms of direct impacts stemming from THz systems, high dimensional array data poses complexity challenge for signal processing and hardware. This necessitates the use of novel approaches including reduced-rate, sparsity-driven, and learning-based channel estimation and beamforming techniques. 
		
				\item 	Furthermore, distance-dependent bandwidth and Doppler shift should be considered for wideband THz S\&C applications. Transmission range is also critical for accurately estimating the THz channel, which may necessitate novel algorithms based on spherical propagation model. 
		
		\item In order to compensate for beam-split, hardware-based approaches (e.g., TTDs) may be costly for large arrays, and signal processing-based techniques are of great interest.

		\item   Beside the conventional IRS, the simultaneously transmitting and reflecting intelligent surface (STARS)-assisted ISAC provides full-space coverage and more  DoF, hence, opens new research opportunities. 
		
		\item Combining RL and FL can lead to improved S\&C performance while lowering CO and complexity of labeling very large THz array data.
		
		\item Another possible research direction may include the THz-ISAC design in low earth orbit (LEO) satellites which employ hybrid beamformers.

	\end{itemize}
	
}}

	\balance
	\bibliographystyle{IEEEtran}
	\bibliography{IEEEabrv,references_098}

%
%
%
%
%
	


	%

\end{document}